\documentclass[prd,tightenlines, 
nofootinbib,showpacs,preprintnumbers,superscriptaddress]{revtex4}
\usepackage{amsfonts,amsmath,amssymb,amsthm,bbm}
\usepackage{graphicx} 
\usepackage{color}

\newcommand{\be}{\begin{equation}}
\newcommand{\ee}{\end{equation}}
\newcommand{\beq}{\begin{eqnarray}}
\newcommand{\eeq}{\end{eqnarray}}

\begin{document}

\title{Creation of entangled universes avoids the big bang singularity}
\author{Salvador J. Robles-P\'{e}rez}
\affiliation{Instituto de F\'{\i}sica Fundamental, Consejo Superior de Investigaciones Cient\'{\i}ficas, Serrano 121, 28006-Madrid, Spain.}
\affiliation{Estaci\'{o}n Ecol\'{o}gica de Biocosmolog\'{\i}a, Pedro de Alvarado, 14, 06411-Medell\'{\i}n, Spain.}
\date{\today}

\begin{abstract}
The creation of universes in entangled pairs may avoid the initial singularity and it would have observable consequences in a large macroscopic universe like ours, at least in principle. In this paper we describe the creation of an entangled pair of universes from a double instanton, which avoids the initial singularity, in the case of a homogeneous and isotropic universe with a conformally coupled massless scalar field. The thermodynamical properties of inter-universal entanglement might have observable consequences on the properties of our single universe provided that the thermodynamics of entanglement is eventually related to the classical formulation of thermodynamics.
\end{abstract}

\maketitle



\section{Introduction}

From the very beginning of human knowledge the creation of the universe has been one of the most exciting, fundamental, and intriguing questions of natural philosophy. In contemporary science, it has been the prime feature of quantum cosmology and it is profoundly related to the existence of space-time singularities and the need of a quantum theory of gravity that would presumably explain or avoid them.

In quantum cosmology, the birth of the universe is deeply related to the boundary conditions that we impose on the state of the universe. Different boundary conditions have been proposed in the literature \cite{Hawking1982, Hartle1983, Vilenkin1982, Vilenkin1986, Vilenkin1994, Gott1998, Penrose1979, Hawking1996, Linde1986}. Among them, two main proposals have become customary in quantum cosmology: the Hartle-Hawking's no boundary proposal \cite{Hawking1982, Hartle1983, Hawking1983} and the Vilenkin's tunneling proposal \cite{Vilenkin1982, Vilenkin1984, Vilenkin1986}. In both cases, the universe is said to be created from \emph{nothing}, where by \emph{nothing} we should not understand the absolute meaning of nothing, i.e., something to which we can ascribe no properties, but rather a classically forbidden region of the space-time where space, matter, and above all time, do not physically exist as such.

The boundary conditions to be imposed on the state of the universe have usually been considered of metaphysical nature and its choice thus a sort of  taste. Some particular proposal \cite{Vilenkin1994, Linde1990} has  claimed to be preferable in order to have a suitable long enough inflationary stage of the universe that would explain the observed homogeneity and isotropy. However, counter-arguments have also been given \cite{Barvinsky1990, Barvinsky1994, Hartle2008, Hartle2008b}. The main problem for an observational choice of the boundary conditions of the universe is that the rapid expansion of the inflationary period would erase any trace of the pre-inflationary stage of the universe \cite{Linde1990}. Besides, the question would be more unsettled than ever if the results of the Planck mission \cite{Ade2013a, Ade2013b} eventually disfavor typical models of inflation \cite{Ijjas2013}. It would be therefore extremely interesting if any proposal for the creation of the universe could provide us with testable effects on a large macroscopic universe like ours.

The multiverse \cite{Tegmark2003, Carr2007, Mersini2008b} is, on the other hand, a new scenario where the creation of the universe can now be depicted. It is worth noticing that the multiverse entails a whole new paradigm that changes many of the preconceptions made in the cosmology of the twentieth century. In particular, in the context of a multiverse the creation of universes could be given in entangled pairs. We shall show in this paper that the creation of universes in entangled pairs may avoid the initial singularity because the universes never shrink to the singular hypersurface of vanishing volume. Furthermore, for an observer inside one of the universes of the entangled pair, the universe would appear to be in a thermal state whose properties would depend on the properties of entanglement \cite{RP2012b, RP2012c}. Thus, the creation of universes in entangled pairs and the underlying boundary conditions might have observable consequences in the thermodynamical properties of one single universe provided that the thermodynamics of entanglement is eventually related to the customary formulation of thermodynamics, as it is expected \cite{Vedral2002, Anders2007, Brandao2008, Amico2008}.

The paper is outlined as follows. In Sec. II it is revisited the creation of a universe from nothing. Following the arguments of Gott \cite{Gott1998} and Barvinsky \cite{Barvinsky2006, Barvinsky2007a, Barvinsky2007b}, we conclude that the creation of the universe from \emph{nothing} is not a realistic scenario. In Sec. III it is described the creation of universes within a multiverse scenario either from a preexisting baby universe or in entangled pairs from a double instanton that avoids the initial singularity. We then make some comments on the thermodynamical properties of entanglement in the so-called Barvinsky-Gott multiverse. In Sec. IV we draw some comment on the boundary conditions to be imposed on the state of the whole multiverse and we finally summarize and conclude in Sec. V.


\section{Creation of the universe from nothing revisited}

Let us review the customary picture of the creation from \emph{nothing} of a de-Sitter universe \cite{Vilenkin1982, Vilenkin1984, Hartle1983, Hawking1984, Kiefer2007}, with a value $\Lambda$ of the cosmological constant. Let us also consider a conformally coupled massless scalar field which, consistently with the geometry, is homogeneous, $\varphi = \varphi(t)$. The Wheeler-DeWitt equation separates \cite{Gott1998}
\begin{eqnarray}\label{eq1}
\frac{1}{2} \left( -\frac{d^2}{d\chi^2} + \chi^2 \right) \phi(\chi) = E \phi(\chi) , \\ \label{eq2}
\frac{1}{2} \left[ -\frac{1}{a^p} \frac{d}{da}\left( a^p  \frac{d}{da} \right) +  \left( a^2 - \frac{\Lambda}{3} a^4 \right) \right] \psi(a) = E \psi(a) ,
\end{eqnarray}
where $\Psi(\chi, a) \equiv \phi(\chi)\psi(a)$ is the wave function of the universe, with $\chi \equiv (4\pi/3)^{1/2} \varphi a$, $a$ is the scale factor, $E\equiv E_n =(n+\frac{1}{2})$ is the energy level of the scalar field, and $p$ is a constant determining the operator ordering \cite{Hartle1983, Gott1998}. The first of these equations is the equation of a quantum harmonic oscillator, which can be solved in terms of Hermite polynomials, and the second equation can formally be written as the 'classical' equation of a harmonic oscillator
\begin{equation}\label{WDW2}
\ddot{\psi}(a) + \frac{\dot{\mathcal{M}}}{\mathcal{M}} \dot{\psi}_n(a) + \omega^2_n \psi_n(a) = 0 ,
\end{equation}
with a scale factor dependent \emph{mass} and \emph{frequency} given by, $\mathcal{M}\equiv\mathcal{M}(a) = a^p$ and 
\begin{equation}\label{frequency}
\omega_n\equiv\omega_n(a) = \sqrt{\frac{\Lambda}{3} a^4 - a^2 + 2 E_n} ,
\end{equation}
respectively, where the scale factor formally plays the role of the time variable in Eq. (\ref{WDW2}). For values of the scale factor well above from the Planck scale we can use the WKB approximation, 
\begin{equation}
\psi_n(a) \approx \sqrt{\frac{\hbar}{2 \mathcal{M}(a) \omega_n(a)}} e^{\pm\frac{i}{\hbar} S_n(a)} ,
\end{equation}
where
\begin{equation}
S_n(a) = \int^a da' \, \omega_n(a') .
\end{equation}
The frequency $\omega_n(a)$, given by Eq. (\ref{frequency}), can also be written as
\begin{equation}
\omega_n(a) = H \sqrt{(a^2 -a_+^2) (a^2-a_-^2)} ,
\end{equation}
with
\begin{equation}\label{amasmenos}
a_\pm^2 \equiv \frac{1}{2 H^2} \left( 1 \pm (1- 4 C H^2)^\frac{1}{2} \right) ,
\end{equation}
and $H^2 \equiv \Lambda$, $C \equiv 2 E_n = 2n + 1$.

\begin{figure}
\centering
\includegraphics[width=8cm]{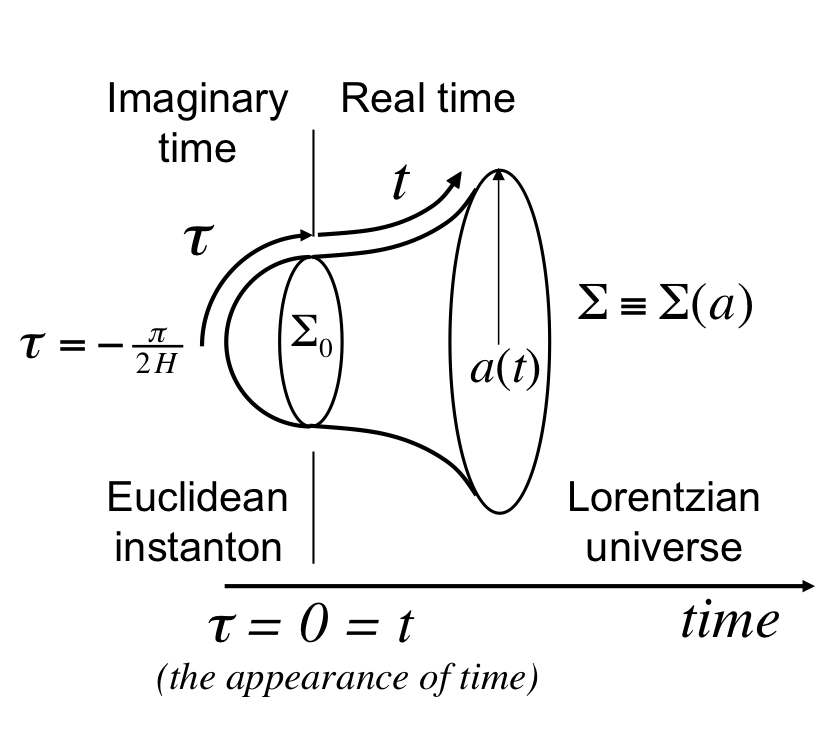}
\caption{The creation of a de-Sitter universe from \emph{nothing} (i.e., from a de-Sitter instanton).}
\label{fig01}
\end{figure}

\begin{figure}
\centering
\includegraphics[width=8cm]{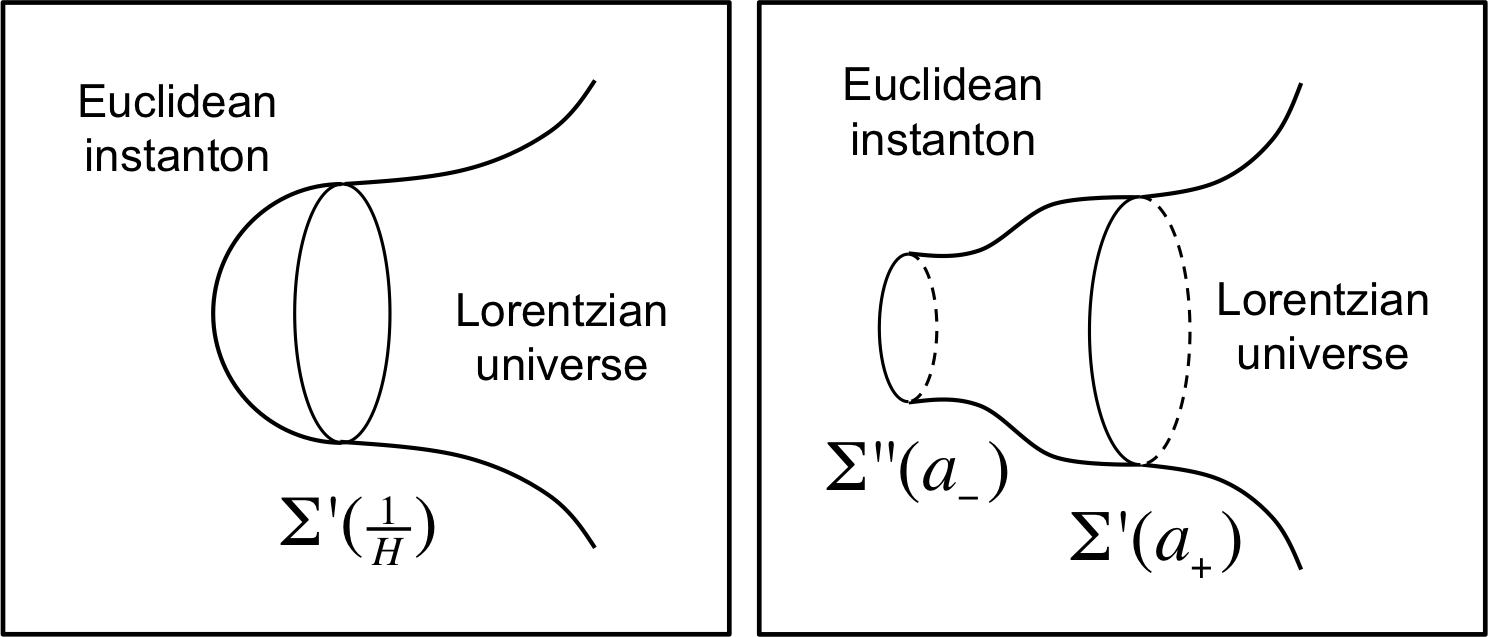}
\caption{Euclidean instantons for the cases: (a) $C=0$ ($a_-=0$, $a_+ = \frac{1}{H}$) and, (b) $C\neq 0$.}
\label{fig01b}
\end{figure}

Let us first consider the scenario that corresponds to the value $C = 0$ \cite{Hartle1983}. Then, $a_+ = \frac{1}{H}$ and $a_- = 0$, and there are thus two regions (see Figs. \ref{fig01}-\ref{fig01b}a). For values $a > a_+$ the Friedmann equation,
\begin{equation}\label{Feq}
\frac{\partial a}{\partial t} = \frac{\omega_n(a)}{a} ,
\end{equation}
has real solution given by
\begin{equation}\label{Lsol}
a(t) = \frac{1}{H} \cosh H t ,
\end{equation}
with $t \in [0, \infty)$, that represents an exponentially expanding universe at late times. For values $a < a_+$, the Friedmann equation (\ref{Feq}) has no real solution. However, we can perform a Wick rotation to Euclidean time, $t \rightarrow i \tau$, by means of which the solution of the transformed Friedmann equation is the analytical continuation of Eq. (\ref{Lsol}), i.e.,
\begin{equation}\label{Esol}
a_E(\tau) = \frac{1}{H} \cos H \tau ,
\end{equation}
with, $\tau \in (-\frac{\pi}{2 H}, 0)$. The solution (\ref{Esol}) represents the Euclidean scale factor of a de-Sitter instanton that shrinks to zero at the Euclidean time $\tau = -\frac{\pi}{2 H}$. The Euclidean scale factor grows then from the value $a_E=0$ to the value $a_E = \frac{1}{H}$, at the Euclidean time $\tau = 0$, where it finds the Lorentzian region and appears there as a universe being created from \emph{nothing} with a length scale of $a_0 \equiv \frac{1}{H}$. That is the customary picture of a de-Sitter universe created from nothing \cite{Vilenkin1982, Hawking1984, Kiefer2007}, which is depicted in Fig. \ref{fig01}.

However, the value $C=0$ in Eq. (\ref{amasmenos}) might be unrealistic. Let us notice that $C \equiv 2 n + 1 = 1$ for the ground state of the scalar field, i.e., for $n=0$, and thus the value $C= 0$ clearly violates the uncertainty principle \cite{Gott1998}. Indeed, the vacuum fluctuations of the ground state of the matter fields prevent the de-Sitter instanton to collapse making the quantum state of the universe be given by a mixed state rather than by a pure state \cite{Barvinsky2006}. In Ref. \cite{Hartle1983}, Hartle and Hawking argued that the renormalization of the matter fields might cancel out the zero-point energy. However, as it is pointed out in Ref. \cite{Gott1998}, there is no reason to expect such an exact cancellation. In fact, in Ref. \cite{Barvinsky2006} it is computed the renormalization of the conformally coupled scalar field considered in Ref. \cite{Hartle1983} and not only it does not cancel out the zero-point energy but it prevents it from being zero, actually. It turns out to be a remarkable outcome because it shows that the customary formulation of a universe being created from \emph{nothing} would not be a realistic scenario and the consideration of the three regions depicted in Fig. \ref{fig02} becomes unavoidable.

For values of $C$ different from zero, the picture significantly departures from the scenario given above. For the value, $1 > 4 C H^2 > 0$ in Eq. (\ref{amasmenos}), there is one classically forbidden region placed between two allowed regions, labelled as $II$, $I$, and $III$, respectively, in Fig. \ref{fig02}. We are then  just left with two possibilities for the quantum creation of a universe in region $III$: either it is created by a tunneling process from a preexisting baby universe of region $I$ \cite{Gott1998} (see, Fig. \ref{fig02}), or it is created from a double instanton \cite{Barvinsky2006} that would eventually give rise to an entangled pair of universes (see, Fig. \ref{fig03}).

\section{Creation of universes in the multiverse}

\subsection{Creation of the universe from \emph{something}}

\begin{figure}
\centering
\includegraphics[width=8cm]{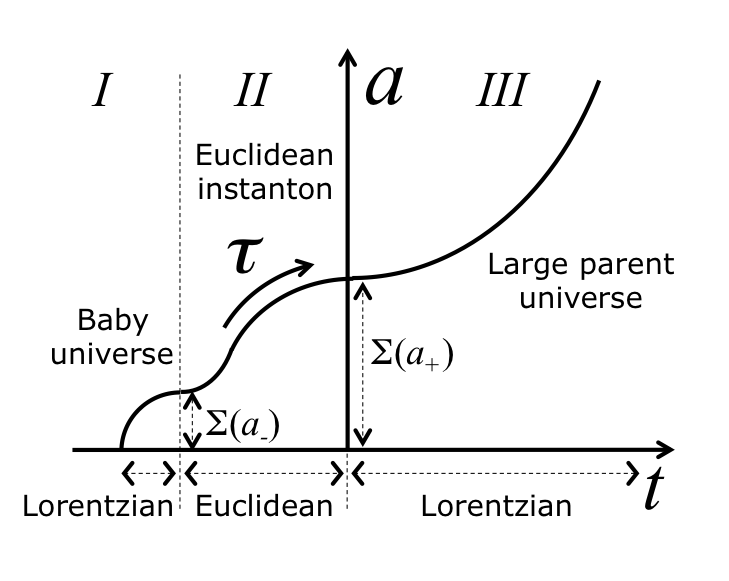}
\caption{The creation of a large parent universe from a baby universe.}
\label{fig02}
\end{figure}

Let us first describe the regions labelled by $I$, $II$, and $III$, in Fig. \ref{fig02}. In the region $I$, the solution of the Friedman equation (\ref{Feq}) is given by
\begin{equation}
a_I(t) = \frac{1}{\sqrt{2}H} \sqrt{1-(1- 4C H^2)^\frac{1}{2} \cosh 2Ht} ,
\end{equation}
where $t_* \geq t \geq - t_*$, with $t_* = \frac{1}{2 H} {\rm arcosh} \frac{1}{\sqrt{1-4CH^2}}$. It describes an oscillating universe that for sufficiently small values of the scale factor, or sufficiently small values of the cosmological constant (i.e. $H\ll 1$), behaves like a radiation dominated universe with
\begin{equation}
a_I(t) \approx \sqrt{C - t^2} ,
\end{equation}
with, $t_* \approx \sqrt{C}$. At $t=0$ the scale factor reaches its maximum value, $a_-$, and it contracts afterwards until it reaches the value $a_I = 0$, at $t=t_*$. At the moment of maximum expansion $a=a_-$ and the universe can tunnel out to region $III$, where
\begin{equation}
a_{III}(t) = \frac{1}{\sqrt{2}H} \sqrt{1+ (1- 4C H^2)^\frac{1}{2} \cosh 2Ht} ,
\end{equation}
with $t \in [0,\infty)$. The scale factor $a_{III}$ describes a universe that starts out from the value $a_{III}=a_+$, at $t=0$, and eventually expands like a vacuum dominated universe, with $a_{III} \approx \frac{1}{2 H} e^{Ht}$. Finally, the classically forbidden region $II$ can be described by an Euclidean instanton with Euclidean scale factor given by \cite{Gott1998, Barvinsky2006}
\begin{equation}
a_{II}^E(\tau) = \frac{1}{\sqrt{2}H} \sqrt{1-(1- 4C H^2)^\frac{1}{2} \cos 2H\tau} ,
\end{equation}
where $\tau \in [0,\frac{\pi}{2 H}]$ is the Euclidean time, for which $a_{II}^E \in [a_-, a_+]$. The picture is then the following: a large vacuum-dominated universe is created through a quantum tunneling process from a pre-existing baby universe which oscillates between a minimal and a maximal value of its length scale. At the maximum value of its scale factor, $a=a_-$, the probability to tunnel out to region $III$ with a value of the scale factor given by $a=a_+$ is given by
\begin{equation}
P \propto e^{-I} ,
\end{equation}
with \cite{Alonso2013}
\begin{eqnarray}
I &=& H \int_{a_-}^{a_+} da \, \sqrt{(a^2 - a_-^2)(a_+^2 - a^2)} \\
&=& \frac{a_+ H}{3} \left[ \left( a_+^2 + a_-^2 \right) E(q) - 2 a_-^2 K(q) \right] ,
\end{eqnarray}
where $K(q)$ and $E(q)$ are the complete elliptical integrals of first and second kind, respectively, with $q \equiv \sqrt{1-\frac{a_-^2}{a_+^2}}$.

Let us notice that the baby universes are expected to exist in the space-time background of a parent universe because they are usually considered to be the quantum fluctuations of the gravitational vacuum of a larger universe \cite{Strominger1990}. If the universes are thus created from other preexisting universes, then, the most appropriate boundary condition for the universe, and for the whole multiverse itself, seems to be that proposed by Gott that the universe is its own mother \cite{Gott1998}. In such scenario, there are two differentiated regions in the multiverse: one populated by baby universes of Planck length (or slightly greater), in which closed temporal curves (CTC's) are allowed to exist, and another region populated with large macroscopic universes which remain classically disconnected.  The causal relations between events in a macroscopic universe are preserved because CTC's are not allowed therein. However, baby universes would pop up as quantum fluctuations of the gravitational vacuum of the macroscopic universes. These newly created baby universes live in the region of Planck scale where CTC's are allowed and, therefore, some of them might be joined by means of a CTC to the baby universe from which the macroscopic universe came \emph{initially} up. The universe is thus its own mother \cite{Gott1998} and the corresponding closed self-generated multiverse turns out to be then a multiply connected set of space-time regions being created from themselves, where terms like \emph{before} or \emph{later}, \emph{initial} or \emph{final} lose their absolute meaning, which would be just restricted to the local space-time region of the speaker who uses them.


\begin{figure}
\centering
\includegraphics[width=10cm]{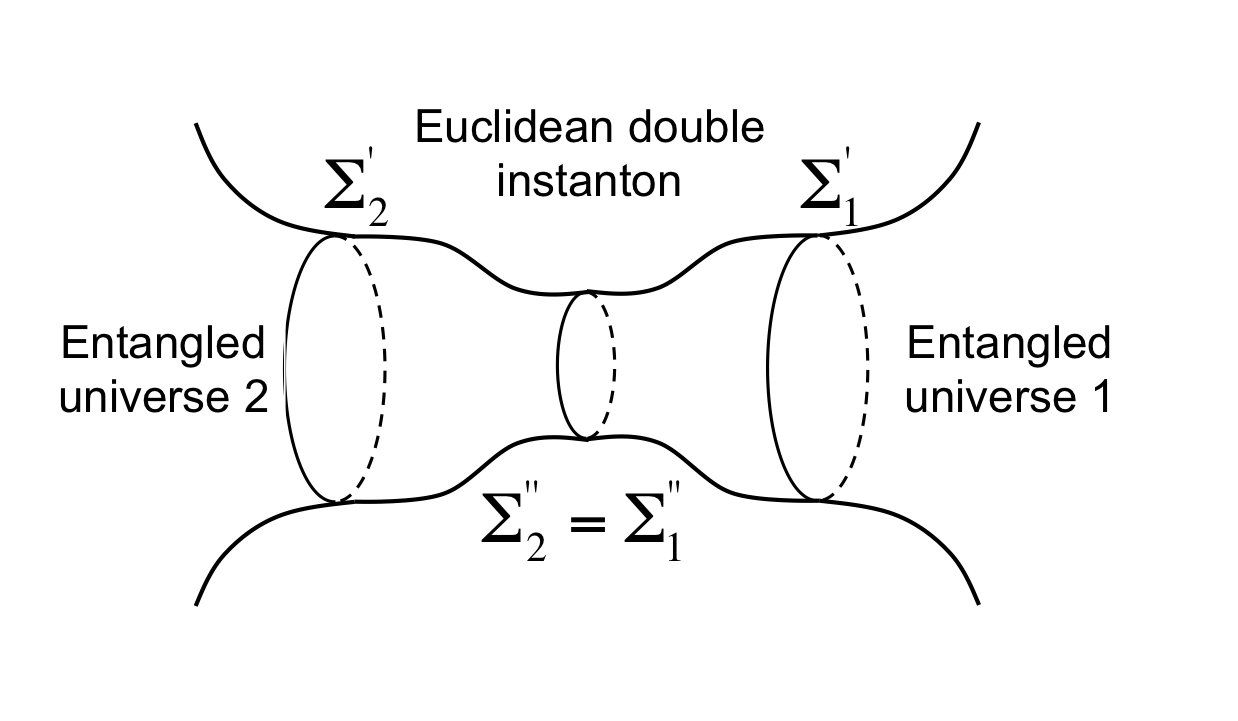}
\caption{The creation of a pair of entangled universes from a Euclidean double instanton.}
\label{fig03}
\end{figure}

\subsection{Creation from nothing of an entangled pair of universes}

\begin{figure}
\centering
\includegraphics[width=8cm]{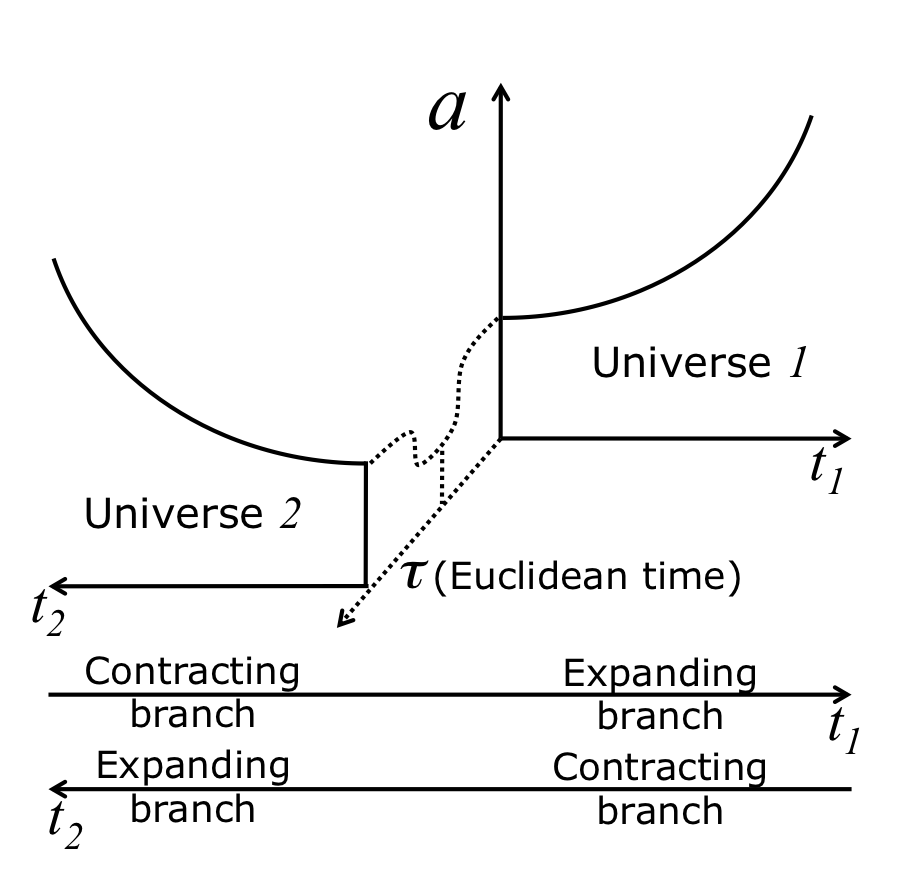}
\caption{Expanding and contracting branches in a pair of entangled universes.}
\label{fig04}
\end{figure}

There is still another possibility for the universes to be created from nothing, i.e., from the Euclidean region, with no need therefore of a former, preexisting universe. However, as it has been previously argued, the universes cannot  be created from nothing individually but they have to come up in entangled pairs which would be the analytic continuation of a double Euclidean instanton like the one depicted in Fig. \ref{fig03}. Let us notice that two Euclidean instantons, like those depicted in Fig. \ref{fig01b}b and Fig. \ref{fig03}, can be joined by matching their hypersurfaces, $\Sigma''_1$ and $\Sigma''_2$, evaluated at $a_-$. The value of $a_-$ in Eq. (\ref{amasmenos}) depends on the value $n$ of the mode of the conformal scalar field so the two instantons can only be joined for the same mode of their scalar fields, with a composite quantum state of the multiverse given by
\begin{equation}\label{comS}
| \Psi_{1,2}\rangle = \sum_n \alpha_n | 0_1^n 0_2^n \rangle + \beta_n | 1_1^n 1_2^n \rangle   ,
\end{equation}
with appropriate coefficients, $\alpha_n$ and $\beta_n$, that depend on the boundary conditions imposed on the state of the whole multiverse, and where $| 0_1^n\rangle$ ($| 0_2^n\rangle$) and $| 1_1^n\rangle$ ($| 1_2^n\rangle$) are the zero-universe and one-universe states, respectively, of the universe $1$($2$) for a value $n$ of the mode of their scalar fields. The composite state (\ref{comS}) thus represents the state of a multiverse made up of entangled pairs of universes which are either created in entangled pairs or not created at all. From the point of view of an observer inhabiting the expanding branch of the universe, the partner universe of the entangled pair can be seen as the corresponding contracting branch. However, for the inhabitant of the partner universe, her universe is expanding as well. This would be just the effect of having the opposite directed time variables (see, Fig. \ref{fig04}), an effect that seems to be similar to that proposed in Ref. \cite{Linde1988}. For each observer, joining the contracting branch to the expanding branch would globally appear as a universe with a \emph{bounce}.

Regarding the flux of particles provided by the Euclidean sector, we should notice that the particles injected into each Lorentzian universe would appear to have an overall different sign of the energy with respect to each other, from the point of view of internal observers, because the reversal correlation of their time variables (see, Fig. \ref{fig04}). However, the overall sign of the energy of particles is a matter of convention \cite{Linde1988} provided that there are no particles with the opposite sign of the energy in the same universe. As it happens with the antipodal symmetry proposed by Linde in Ref. \cite{Linde1988}, there would be no problem here either because particles belonging to different universes do not interact with each other due to the Euclidean gap. Let us also notice that the two universes of the entangled pair are otherwise similar to those created individually except for the thermodynamical properties of entanglement. Therefore, once the (real) cosmic time has appeared the light cones behave as they usually do in a de Sitter like universe.

The Fock space of number states used in Eq. (\ref{comS}) is well-defined, in an appropriate representation. The general solution $\Psi(a,\chi)$ of Eqs. (\ref{eq1}-\ref{eq2}) can generally be written as
\begin{equation}\label{qsmultiverse}
\Psi(a,\chi) = \sum_n c_n \phi_n(\chi) \psi_n(a) + c.c. ,
\end{equation}
where, $c.c.$ means the complex conjugated and $c_n$ is a constant. The third quantization formalism consists of promoting the wave function $\Psi(a,\chi)$ to an operator much in the same way as it is done in a quantum field theory. Then, the constants $c_n$ and $c_n^*$ turn out to be, in an appropriate representation, the creation and annihilation operators of universes, respectively, and the eigenstates of the corresponding number operator, $\hat{N} \equiv \hat{c}_n^\dag \hat{c}_n$, would represent the number of universes in the multiverse. However, it is not easy at all to determine which representation accurately describes the number of universes in the multiverse. Our proposal \cite{RP2012, Garay2013} is that the number of universes of the multiverse should be independent of the value of the scale factor of a particular single universe and, then, the appropriate representation would be an invariant representation (see, Sec. IV).

Furthermore, the wave functions $\Psi_n(a,\chi)\equiv \phi_n(\chi)\psi_n(a)$ form a basis of the space of solutions of the Wheeler-DeWitt equation which is orthonormal under the scalar product
\begin{equation}
(\Psi_n, \Psi_m) \equiv - i \int_{-\infty}^\infty d\chi \, W^{-1}(a) \left( \Psi_n \frac{\partial}{\partial a} \Psi_m^* - \Psi_m^* \frac{\partial}{\partial a} \Psi_n \right) ,
\end{equation}
where $W(a)$ is the Wronskian of Eq. (\ref{WDW2}), provided than the wave functions $\psi_n(a)$ satisfy the normalization condition
\begin{equation}
W^{-1} ( \psi_n \frac{\partial}{\partial a} \psi_m^* - \psi_m^* \frac{\partial}{\partial a} \psi_n ) = i .
\end{equation}
Then, the usual orthonormality conditions \cite{Birrell1982, Mukhanov2007}
\begin{equation}\label{orthonormality}
(\Psi_n, \Psi_m) = \delta_{nm} , \,\, (\Psi_n^*, \Psi_m^*) = - \delta_{nm} , \,\,  (\Psi_n, \Psi_m^*) = 0 ,
\end{equation}
are satisfied. The quantum solution of the wave function of the universe is then well-defined for the whole range where the approximations made in the model are valid and we can define a Fock space for the solutions of the wave function of the multiverse.

Let us also remark that there is no initial singularity in an entangled pair of universes created from \emph{nothing} because the universes never reach the singular hypersurface  of vanishing volume, located at $a=0$. The minimum value of their scale factors is in fact given by $a_-$, which can be of some orders above the Planck scale \cite{Garay2013}. For instance, let us consider that $\Lambda$ is the effective value of the potential of another field, $\xi$, with mass $m$ that undergoes a slow-roll regime, i.e., $\Lambda \equiv V(\xi_0)$. The universes would be then continuously created in an eternal self-inflationary process \cite{Linde1986, Linde1990} with values $M_P^4 \gtrsim V(\xi_0) \gtrsim \lambda M_P^4$, where $\lambda \equiv \frac{m}{M_P}$. For a realistic inflationary scenario \cite{Linde1990} of $m \approx 10^{12} {\rm GeV}$, $\lambda \sim 10^{-7}$ and the length $l \sim H^{-1}$ of the universes being created can then be as large as \cite{Linde1990, Garay2013} $l \sim \lambda^{-\frac{1}{2}} l_P \sim 10^3 l_P$. That means that $a_+$ in Eq. (\ref{amasmenos}) has to be of the same order, i.e., $a_+ \sim 10^3 l_P$. There is still room therefore for a value $a_-$ that satisfies $a_+ \gtrsim a_- \gg l_P$, avoiding thus the singular region of the Planck scale and bellow.

On the other hand, the inter-universal entanglement might have observable consequences on the dynamical and thermodynamical properties of one single universe of the multiverse \cite{RP2012, RP2012b, RP2012c, RP2012d}. Let us notice that the state (\ref{comS}) represents a composite pure entangled state that in the density matrix formalism reads
\begin{equation}
\rho \equiv |\Psi_{1,2} \rangle \langle \Psi_{1,2} | .
\end{equation}
The reduced density matrix that represents the quantum state of one single universe of the entangled pair can be obtained by tracing out the degrees of freedom of the partner universe, i.e., 
\begin{equation}\label{reducedRho}
\rho_1 \equiv {\rm tr}_{2} \rho = \sum_n |\alpha_n|^2 |0_1^n\rangle \langle 0_1^n| + |\beta_n|^2 |1_1^n\rangle \langle 1_1^n|   ,
\end{equation}
where in general the coefficients $\alpha_n$ and $\beta_n$ depend on the value of the scale factor, $\alpha_n \equiv \alpha_n(a)$ and $\beta_n \equiv \beta_n(a)$. They determine the entanglement properties of the entangled pair of universes. For instance, the entropy of entanglement associated to the reduced density matrix (\ref{reducedRho}), given by the von-Neumann formula, yields
\begin{equation}
S_{ent} = - \sum_n |\alpha_n|^2 \log |\alpha_n|^2 + |\beta_n|^2 \log |\beta_n|^2 .
\end{equation}
It depends therefore on the entanglement properties of the entangled pair of universes. That could have observable consequences on the properties of one single universe of the entangled pair provided that thermodynamics of entanglement can eventually be related to the customary formulation of thermodynamics, as it is expected (see Refs. \cite{Vedral2002, Anders2007, Brandao2008, Amico2008}). In that case, the entropy of entanglement could provide us with an arrow of time that might have consequences on the growth of cosmic structures in the earliest stages of the universe \cite{RP2012}.

In a more general multiverse scenario with inter-universal entangled and squeezed states \cite{RP2012, RP2012c, RP2012d}, the reduced density matrix that describes the state of one single universe turns out to be given by a thermal state whose properties are indistinguishable from a classical mixture. The energy of entanglement between two or more universes might thus provide us with a contribution to the energy density of each single universe and therefore have observable consequences, at least in principle  \cite{RP2012c, RP2012d}.

Furthermore, the global picture of the multiverse become even richer if quantum effects of backreaction and renormalization are taken into account \cite{Barvinsky2006}. The cosmological landscape turns out to be then limited to a bounded range of the cosmological constant, $\Lambda_{max} \geq \Lambda \geq \Lambda_{min}$, where $\Lambda_{max}$ arises from the renormalization of the conformally invariant field and $\Lambda_{min}$ comes from the backreaction \cite{Barvinsky2006}. Moreover, for each single mode of the conformally invariant scalar field it appears a countable sequence of one-parameter families of instantons with continuous subsets, called \cite{Barvinsky2006} \emph{garlands}, for which
$$
C_{max}^{(n)} \geq C^{(n)} \geq C_{min}^{(n)} \,\, , \, \, \Lambda_{max}^{(n)} \geq \Lambda > \Lambda_{min}^{(n)} ,
$$
with \cite{Barvinsky2006} $$\Lambda_{max}^{(n)} - \Lambda_{min}^{(n)} \sim \frac{1}{n^4}.$$Instantons with $\Lambda > \Lambda_{max}$ or $\Lambda < \Lambda_{min}$ are exponentially suppressed \cite{Barvinsky2006}. Then, following a similar mechanicsm to that described previously we can pose, for each single member of the continuous set, bipartite and multi-partite inter-universal entangled states that would be the analytic continuation of the \emph{garlands}. Besides, interacting terms \cite{Alonso2012} can also be posed among the universes of the Barvinsky-Gott landscape \cite{Alonso2013} which could induce quantum transitions among the states that correspond to different values of the parameters of the landscape. The phenomenology of the multiverse becomes then extremely rich, still to be explored.


\section{Comments on the boundary conditions of the multiverse}

The appropriate choice of boundary conditions to be imposed on the state of the whole multiverse and its relationship with the boundary conditions that are customary imposed on the state of a single universe is a matter that deserves further investigation and a deeper understanding of the particular model of the multiverse being considered. It is nevertheless a significant choice because the boundary condition of the multiverse would determine, among other things, the kind of residual correlations that may exist among the universes of the multiverse and thus the possibility of detecting their influence in a single universe like ours. In other words, if we could observe indirect effects of other universes in the properties of our own universe we could then attempt to establish the appropriate boundary conditions of the whole multiverse.

In the context of the multiverse being analyzed in this paper, it seems clear that if the universes are created from pre-existing baby universes, which are considered the quantum fluctuation of the gravitational vacuum of a large parent universe, a natural boundary condition is the Gott-Li's proposal of a universe which is the mother of itself \cite{Gott1998}.

On the other hand, in a multiverse made up of entangled pairs of universes, the boundary condition that the number of universes should not depend on the value of the scale factor of a particular single universe \cite{RP2012b} imposes some restrictions on the boundary conditions of the single universes.  The relationship between the boundary condition of the multiverse and those usually imposed on the state of single universes, i.e., the no-boundary proposal \cite{Hawking1982, Hartle1983, Hawking1983} and the tunneling boundary condition \cite{Vilenkin1982, Vilenkin1984, Vilenkin1986}, can plainly be seen in the model of a multiverse of flat de Sitter universes endorsed with a minimally coupled massless scalar field, where exact analytic solutions can be found \cite{RP2012b}. In that case, the quantum state (\ref{qsmultiverse}) of the multiverse can be written instead as 
\begin{equation}\label{decompositiona}
\hat{\Psi}(a,\varphi) = \int d k e^{i k \varphi} \psi_{k}(a) \hat{c}_{k}^\dag + e^{-i k\varphi} \psi^*_{k}(a) \hat{c}_{k} ,
\end{equation}
where unlike in Eq. (\ref{qsmultiverse}), the variable $k$ runs here over a continuous spectrum (it is the value of the third quantized momentum associated to the scalar field variable, see Ref. \cite{RP2012b}). The amplitudes $\psi_{k}(a)$ satisfy then the Bessel equation
\begin{equation}\label{Bessel}
a^2 \ddot{\psi}_{k} + a\dot{\psi}_{k} + (H^2 a^{6} + k^2) \psi_{k} = 0,
\end{equation}
with, $\dot{\psi}_{k}\equiv \frac{\partial \psi_{k}}{\partial a}$ and $H^2 = \Lambda$. The usual operators of the harmonic oscillator, $\hat{c}_{k}\equiv \sqrt{\frac{\omega_{0k}}{2}} (\hat{\psi} + \frac{i}{\omega_{0k}} \hat{p}_\psi)$ and $\hat{c}_{k}^\dag \equiv \sqrt{\frac{{\omega}_{0k}}{2}} (\hat{\psi} - \frac{i}{{\omega}_{0k}} \hat{p}_\psi)$, with $\omega_{0k} = \sqrt{H^2 a_0^4 + k^2}$, are interpreted in the third quantization formalism as the annihilation and creation operators, respectively, of $k$-modes of a de Sitter universe with a cosmological constant $\Lambda$ and a boundary hypersurface $\Sigma_{0} \equiv \Sigma(a_0)$. The branches of the universe created or annihilated by $\hat{c}_k^\dag$ and $\hat{c}_k$, respectively, eventually depend on the boundary condition that is imposed on the universe. If we impose the Vilenkin's tunneling condition \cite{Vilenkin1986} that the only modes that survive the quantum barrier are the outgoing modes that correspond in the minisuperspace to the expanding branches of the universe, then, the amplitudes $\psi_k$ are given by \cite{RP2012b}
\begin{equation}\label{eq04}
\psi_k(a)= \sqrt{\frac{\pi}{12}} e^{\frac{\pi k}{6}} \mathcal{H}_{\frac{i k}{3}}^{(2)}(\frac{H}{3} a^3) ,
\end{equation}
where $\mathcal{H}_{\nu}^{(2)}(x)$ is the Hankel function of second kind and order $\nu$. The normalization constant in Eq. (\ref{eq04}) has been chosen so that  the orthonormality conditions (\ref{orthonormality}) hold. The modes given in Eq. (\ref{eq04}) correspond to the expanding branches of the universe because in the semiclassical regime (for large values of the scale factor) \cite{Abramovitz1972}
\begin{equation}\label{eq05}
\mathcal{H}_\nu^{(2)}(\frac{H}{3} a^3) \sim a^{-\frac{3}{2}} e^{- i S_c(a) }  ,
\end{equation}
where, $S_c(a) = \frac{H}{3} a^3$, is the classical action.  Then, the momentum operator, which is defined by the equation $\hat{p}_a \psi(a) \equiv - i \hbar \frac{\partial\psi(a)}{\partial a}$, is highly picked around the value of the classical momentum \cite{Halliwell1987}, $p_a^c \equiv - a \frac{\partial a}{\partial t}$, and it then follows that $\frac{\partial a}{\partial t} \approx + \frac{1}{a} \frac{\partial S_c}{\partial a}$, which corresponds to the expanding branch of the Friedmann equation (similarly, the Hankel function of first kind and order $\nu$, $\mathcal{H}_{\nu}^{(1)}(x)$, would correspond in the semiclassical regime to a contracting branch).

If we otherwise impose the no-boundary proposal by requiring regularity conditions at the origin \cite{Halliwell1990}, then, the amplitudes $\psi_k$ are given by \cite{RP2012b}
\begin{equation}\label{eq09}
\bar{\psi}_k = \left( \frac{2 q}{\pi} \sinh \frac{k \pi}{q}  \right)^{-\frac{1}{2}}  \mathcal{J}_{-\frac{i k}{q}} (\frac{\omega_0}{q \hbar} a^q) ,
\end{equation}
where $\mathcal{J}_{\nu}(x)$ is the Bessel function of first kind and order $\nu$. In the semiclassical regime, the ground state corresponds to an equally probable combination of expanding and contracting branches of the universe, as it was originally proposed \cite{Hawking1985}. It is worth noticing that the two sets of modes are related by the Bogoliubov transformation \cite{RP2012b} and that the vacuum state obtained with the no-boundary condition turns out to be a thermal state in the representation of the tunneling wave functions, with a temperature given by, $T = \frac{3}{2 \pi}$, in appropriate units  (see, Ref. \cite{RP2012b}). 

However, it is expected that the appropriate representation of universes in the multiverse would be an invariant representation \cite{RP2010,RP2012d} because then, the number of universes would not depend on the value of the scale factor of a particular single universe. It is worth noticing the important relation in the third quantization formalism between the boundary condition of the multiverse and the representation of universes. Different invariant representations can be chosen (see Ref. \cite{Kim2001} for the invariant representation of a harmonic oscillator). For instance, the Lewis representation \cite{Lewis1969} can be defined in the multiverse as \cite{RP2010,RP2012b}
\begin{eqnarray}\label{b1}
\hat{b}_k(a) &=& \sqrt{\frac{1}{2 }} \left( \frac{\hat{\psi}_k}{R_k} - i (R_k \hat{p}_{\psi_k} - \dot{R}_k \hat{\psi}_k) \right) , \\ \label{b2}
\hat{b}_k^\dag(a) &=& \sqrt{\frac{1}{2 }} \left( \frac{\hat{\psi}_k}{R_k} + i (R_k \hat{p}_{\psi_k} - \dot{R}_k \hat{\psi}_k) \right) ,
\end{eqnarray}
where, $R_k = R_k(a) \equiv \sqrt{\psi_a^2(a) + \psi_b^2(a)}$, being $\psi_a$ and $\psi_b$ two linearly independent solutions of Eq. (\ref{Bessel}) that make real the function $R$. The representation given by the operators (\ref{b1}-\ref{b2}) conserves the number of universes in the multiverse, i.e. $b^\dag b |N,a\rangle = N |N,a\rangle$, with $N\neq N(a)$, and it thus seems to be an appropriate representation for the number of universes in the multiverse.

In terms of the creation and annihilation operators of the Lewis representation, the decomposition (\ref{decompositiona}) of the wave function of the multiverse is given instead by 
\begin{equation}\label{decomposition2}
\Psi(a,\varphi) = \int d k B_k(\varphi, a) \hat{b}_k + B_k^*(\varphi,a) \hat{b}_k^\dag ,
\end{equation}
with, $B_k(\varphi, a) = e^{i k\varphi} \psi_k(a) \mu^* - e^{-i k \varphi} \psi_k^*(a) \nu^*$, where, $\mu\equiv \mu_k(a)$ and $\nu\equiv \nu_k(a)$, are the squeezing parameters that relate the Lewis operators (\ref{b1}-\ref{b2}) to the constant operators $\hat{c}_n$ and $\hat{c}_n^\dag$, i.e. 
\begin{eqnarray}
\hat{b}_k &=& \mu \hat{c}_k + \nu \hat{c}_k^\dag , \\
\hat{b}_k^\dag &=& \mu^* \hat{c}_k^\dag + \nu^* \hat{c}_k ,
\end{eqnarray}
with
\begin{equation}
\mu = \frac{1}{2 \sqrt{\omega_{0 k}}} \left( \frac{1}{R_k} +  R_k \omega_{0 k} + i \dot{R}_k \right) \; \; , \; \; \nu = \frac{1}{2 \sqrt{\omega_{0 k}}} \left( \frac{1}{R_k} -  R_k \omega_{0 k} + i \dot{R}_k \right)
\end{equation}
and $|\mu|^2 - |\nu|^2 = 1$. Thus, the Lewis operators given by Eqs. (\ref{b1}-\ref{b2}) do not represent the creation and annihilation of either isolated or equally probable expanding and contracting branches of the universe but in general a scale factor dependent combination of expanding and contracting branches of the universe. It is an example of how the boundary condition imposed on the state of the multiverse can determine the boundary conditions to be imposed on the single universes. However, different invariant representations can be chosen and besides the choice of the boundary conditions of the multiverse is still far from being settled.


\section{Conclusions}

We have presented a mechanism for which the creation of the universe avoids the initial singularity. First, we have revisited the customary proposal of a universe being created from \emph{nothing}, where by \emph{nothing} we mean the classically forbidden regime of the gravitational field, i.e., the Euclidean region. Following the arguments of Gott \cite{Gott1998} and Barvinsky \cite{Barvinsky2006} we have concluded that it is an unrealistic scenario. The same conclusion can be obtained as well using other models for the scalar field \cite{RP2012b, Garay2013}.

We are then left with just two possibilities for the quantum creation of universes: either the universe is created from a preexisting baby universe or the universes are created from \emph{nothing} in entangled pairs. In the case of a universe being created through a tunneling process from a preexisting universe the most appropriate boundary condition for the universe, and for the multiverse itself, is Gott-Li's boundary condition that the universe is the mother of itself. In the case of the quantum creation of entangled pairs of universes the initial singularity is avoided because the double instanton never reaches the singular hypersurface of vanishing volume, located at $a=0$. The boundary condition that the number of universes of the multiverse would not depend on the value of the scale factor of a particular single universe induces the creation of entangled pairs of universes. The dynamical and thermodynamical properties of one of the universes of the entangled pair would depend on the properties of inter-universal entanglement provided that the thermodynamics of entanglement are eventually related to the customary formulation of thermodynamics.

Finally, let us emphasise that the Barvkinsky-Gott multiverse turns out to be an extremely rich structure made up of entangled sets of universes, associated to each value of the parameters of the landscape, which would be the analytic continuation of the garlands \cite{Barvinsky2006}. Interacting terms among the universes of the Barvinsky-Gott landscape would add a new phenomenology that might have observable consequences on the properties of our single universe, providing us therefore with a prolific new scenario still to be explored.



\begin{thebibliography}{49}
\expandafter\ifx\csname natexlab\endcsname\relax\def\natexlab#1{#1}\fi
\expandafter\ifx\csname bibnamefont\endcsname\relax
  \def\bibnamefont#1{#1}\fi
\expandafter\ifx\csname bibfnamefont\endcsname\relax
  \def\bibfnamefont#1{#1}\fi
\expandafter\ifx\csname citenamefont\endcsname\relax
  \def\citenamefont#1{#1}\fi
\expandafter\ifx\csname url\endcsname\relax
  \def\url#1{\texttt{#1}}\fi
\expandafter\ifx\csname urlprefix\endcsname\relax\def\urlprefix{URL }\fi
\providecommand{\bibinfo}[2]{#2}
\providecommand{\eprint}[2][]{\url{#2}}

\bibitem[{\citenamefont{Hawking}(1982)}]{Hawking1982}
\bibinfo{author}{\bibfnamefont{S.~W.} \bibnamefont{Hawking}},
  \bibinfo{journal}{Astrophysical Cosmology, 563-72. Vatican City: Pontificia
  Academiae Scientarium}  (\bibinfo{year}{1982}).

\bibitem[{\citenamefont{Hartle and Hawking}(1983)}]{Hartle1983}
\bibinfo{author}{\bibfnamefont{J.~B.} \bibnamefont{Hartle}} \bibnamefont{and}
  \bibinfo{author}{\bibfnamefont{S.~W.} \bibnamefont{Hawking}},
  \bibinfo{journal}{Phys. Rev. D} \textbf{\bibinfo{volume}{28}},
  \bibinfo{pages}{2960} (\bibinfo{year}{1983}).

\bibitem[{\citenamefont{Vilenkin}(1982)}]{Vilenkin1982}
\bibinfo{author}{\bibfnamefont{A.}~\bibnamefont{Vilenkin}},
  \bibinfo{journal}{Phys. Lett. B} \textbf{\bibinfo{volume}{117}},
  \bibinfo{pages}{25} (\bibinfo{year}{1982}).

\bibitem[{\citenamefont{Vilenkin}(1986)}]{Vilenkin1986}
\bibinfo{author}{\bibfnamefont{A.}~\bibnamefont{Vilenkin}},
  \bibinfo{journal}{Phys. Rev. D} \textbf{\bibinfo{volume}{33}},
  \bibinfo{pages}{3560} (\bibinfo{year}{1986}).

\bibitem[{\citenamefont{Vilenkin}(1994)}]{Vilenkin1994}
\bibinfo{author}{\bibfnamefont{A.}~\bibnamefont{Vilenkin}},
  \bibinfo{journal}{Phys. Rev. D} \textbf{\bibinfo{volume}{50}},
  \bibinfo{pages}{2581} (\bibinfo{year}{1994}).

\bibitem[{\citenamefont{Gott and Li}(1998)}]{Gott1998}
\bibinfo{author}{\bibfnamefont{J.~R.~I.} \bibnamefont{Gott}} \bibnamefont{and}
  \bibinfo{author}{\bibfnamefont{L.-X.} \bibnamefont{Li}},
  \bibinfo{journal}{Phys. Rev. D} \textbf{\bibinfo{volume}{58}},
  \bibinfo{pages}{023501} (\bibinfo{year}{1998}).

\bibitem[{\citenamefont{Penrose}(1979)}]{Penrose1979}
\bibinfo{author}{\bibfnamefont{R.}~\bibnamefont{Penrose}},
  \emph{\bibinfo{title}{General Relavity: An Einstein Centenary Survey}}
  (\bibinfo{publisher}{Cambridge University Press, Cambridge, UK},
  \bibinfo{year}{1979}), chap.~\bibinfo{chapter}{12}.

\bibitem[{\citenamefont{Hawking and Penrose}(1996)}]{Hawking1996}
\bibinfo{author}{\bibfnamefont{S.~W.} \bibnamefont{Hawking}} \bibnamefont{and}
  \bibinfo{author}{\bibfnamefont{R.}~\bibnamefont{Penrose}},
  \emph{\bibinfo{title}{The Nature of Space and Time}}
  (\bibinfo{publisher}{Princeton University Press, Princeton, USA},
  \bibinfo{year}{1996}).

\bibitem[{\citenamefont{Linde}(1986)}]{Linde1986}
\bibinfo{author}{\bibfnamefont{A.}~\bibnamefont{Linde}},
  \bibinfo{journal}{Phys. Lett. B} \textbf{\bibinfo{volume}{175}},
  \bibinfo{pages}{395} (\bibinfo{year}{1986}).

\bibitem[{\citenamefont{Hawking}(1984{\natexlab{a}})}]{Hawking1983}
\bibinfo{author}{\bibfnamefont{S.~W.} \bibnamefont{Hawking}}, in
  \emph{\bibinfo{booktitle}{Relativity, groups and topology II, Les Houches,
  Session XL, 1983}}, edited by \bibinfo{editor}{\bibfnamefont{B.~S.}
  \bibnamefont{De~Witt}} \bibnamefont{and}
  \bibinfo{editor}{\bibfnamefont{R.}~\bibnamefont{Stora}}
  (\bibinfo{publisher}{Elsevier Science Publishers B. V.},
  \bibinfo{year}{1984}{\natexlab{a}}).

\bibitem[{\citenamefont{Vilenkin}(1984)}]{Vilenkin1984}
\bibinfo{author}{\bibfnamefont{A.}~\bibnamefont{Vilenkin}},
  \bibinfo{journal}{Phys. Rev. D} \textbf{\bibinfo{volume}{30}},
  \bibinfo{pages}{509} (\bibinfo{year}{1984}).

\bibitem[{\citenamefont{Linde}(1990)}]{Linde1990}
\bibinfo{author}{\bibfnamefont{A.}~\bibnamefont{Linde}},
  \emph{\bibinfo{title}{Particle physics and inflationary cosmology}}
  (\bibinfo{publisher}{Harwood academic publishers}, \bibinfo{year}{1990}),
  \eprint{arXiv:hep-th/0503203v1}.

\bibitem[{\citenamefont{Barvinsky and Kamenshchik}(1990)}]{Barvinsky1990}
\bibinfo{author}{\bibfnamefont{A.~O.} \bibnamefont{Barvinsky}}
  \bibnamefont{and} \bibinfo{author}{\bibfnamefont{A.~Y.}
  \bibnamefont{Kamenshchik}}, \bibinfo{journal}{Class. Quant. Grav.}
  \textbf{\bibinfo{volume}{7}}, \bibinfo{pages}{L181} (\bibinfo{year}{1990}).

\bibitem[{\citenamefont{Barvinsky and Kamenshchik}(1994)}]{Barvinsky1994}
\bibinfo{author}{\bibfnamefont{A.~O.} \bibnamefont{Barvinsky}}
  \bibnamefont{and} \bibinfo{author}{\bibfnamefont{A.~Y.}
  \bibnamefont{Kamenshchik}}, \bibinfo{journal}{Phys. Lett. B}
  \textbf{\bibinfo{volume}{332}}, \bibinfo{pages}{270} (\bibinfo{year}{1994}),
  \eprint{gr-qc/9404062}.

\bibitem[{\citenamefont{Hartle et~al.}(2008{\natexlab{a}})\citenamefont{Hartle,
  Hawking, and Hertog}}]{Hartle2008}
\bibinfo{author}{\bibfnamefont{J.~B.} \bibnamefont{Hartle}},
  \bibinfo{author}{\bibfnamefont{S.~W.} \bibnamefont{Hawking}},
  \bibnamefont{and} \bibinfo{author}{\bibfnamefont{T.}~\bibnamefont{Hertog}},
  \bibinfo{journal}{Phys. Rev. D} \textbf{\bibinfo{volume}{77}},
  \bibinfo{pages}{123537} (\bibinfo{year}{2008}{\natexlab{a}}).

\bibitem[{\citenamefont{Hartle et~al.}(2008{\natexlab{b}})\citenamefont{Hartle,
  Hawking, and Hertog}}]{Hartle2008b}
\bibinfo{author}{\bibfnamefont{J.~B.} \bibnamefont{Hartle}},
  \bibinfo{author}{\bibfnamefont{S.~W.} \bibnamefont{Hawking}},
  \bibnamefont{and} \bibinfo{author}{\bibfnamefont{T.}~\bibnamefont{Hertog}},
  \bibinfo{journal}{Phys. Rev. Lett.} \textbf{\bibinfo{volume}{100}},
  \bibinfo{pages}{201301} (\bibinfo{year}{2008}{\natexlab{b}}),
  \eprint{arXiv:0711.4630}.

\bibitem[{\citenamefont{Ade and others
  (Planck~Collaboration)}(2013{\natexlab{a}})}]{Ade2013a}
\bibinfo{author}{\bibfnamefont{P.~A.~R.} \bibnamefont{Ade}} \bibnamefont{and}
  \bibinfo{author}{\bibnamefont{others (Planck~Collaboration)}}
  (\bibinfo{year}{2013}{\natexlab{a}}), \eprint{arXiv:1303.5082}.

\bibitem[{\citenamefont{Ade and others
  (Planck~Collaboration)}(2013{\natexlab{b}})}]{Ade2013b}
\bibinfo{author}{\bibfnamefont{P.~A.~R.} \bibnamefont{Ade}} \bibnamefont{and}
  \bibinfo{author}{\bibnamefont{others (Planck~Collaboration)}}
  (\bibinfo{year}{2013}{\natexlab{b}}), \eprint{arXiv:1303.5076}.

\bibitem[{\citenamefont{Ijjas et~al.}(2013)\citenamefont{Ijjas, Steinhard, and
  Loeb}}]{Ijjas2013}
\bibinfo{author}{\bibfnamefont{A.}~\bibnamefont{Ijjas}},
  \bibinfo{author}{\bibfnamefont{P.~J.} \bibnamefont{Steinhard}},
  \bibnamefont{and} \bibinfo{author}{\bibfnamefont{A.}~\bibnamefont{Loeb}}
  (\bibinfo{year}{2013}), \eprint{arXiv:1304.2785}.

\bibitem[{\citenamefont{Tegmark}(2003)}]{Tegmark2003}
\bibinfo{author}{\bibfnamefont{M.}~\bibnamefont{Tegmark}}, in
  \emph{\bibinfo{booktitle}{Science and ultimate reality: from quantum to
  cosmos}}, edited by \bibinfo{editor}{\bibfnamefont{J.~D.}
  \bibnamefont{Barrow}}, \bibinfo{editor}{\bibfnamefont{P.~C.~W.}
  \bibnamefont{Davies}}, \bibnamefont{and}
  \bibinfo{editor}{\bibfnamefont{C.~L.} \bibnamefont{Harper}}
  (\bibinfo{publisher}{Cambridge University Press, Cambridge, UK},
  \bibinfo{year}{2003}).

\bibitem[{\citenamefont{Carr}(2007)}]{Carr2007}
\bibinfo{editor}{\bibfnamefont{B.}~\bibnamefont{Carr}}, ed.,
  \emph{\bibinfo{title}{Universe or Multiverse}} (\bibinfo{publisher}{Cambridge
  University Press, Cambridge, UK}, \bibinfo{year}{2007}).

\bibitem[{\citenamefont{Mersini-Houghton}(2008)}]{Mersini2008b}
\bibinfo{author}{\bibfnamefont{L.}~\bibnamefont{Mersini-Houghton}}
  (\bibinfo{year}{2008}), \eprint{arXiv:0804.4280v1}.

\bibitem[{\citenamefont{Robles-P{\'e}rez and
  Gonz{\'a}lez-D{\'\i}az}(2013)}]{RP2012b}
\bibinfo{author}{\bibfnamefont{S.}~\bibnamefont{Robles-P{\'e}rez}}
  \bibnamefont{and} \bibinfo{author}{\bibfnamefont{P.~F.}
  \bibnamefont{Gonz{\'a}lez-D{\'\i}az}}, \bibinfo{journal}{(accepted for
  publication in JETP)}  (\bibinfo{year}{2013}), \eprint{arXiv:1111.4128}.

\bibitem[{\citenamefont{Robles-P{\'e}rez}(2012{\natexlab{a}})}]{RP2012c}
\bibinfo{author}{\bibfnamefont{S.~J.} \bibnamefont{Robles-P{\'e}rez}},
  \bibinfo{journal}{Proceedings of the Multiverse and Fundamental Cosmology
  Meeting (Multicosmofun'12)}  (\bibinfo{year}{2012}{\natexlab{a}}),
  \eprint{arXiv:1212.4598}.

\bibitem[{\citenamefont{Vedral and Kashefi}(2002)}]{Vedral2002}
\bibinfo{author}{\bibfnamefont{V.}~\bibnamefont{Vedral}} \bibnamefont{and}
  \bibinfo{author}{\bibfnamefont{E.}~\bibnamefont{Kashefi}},
  \bibinfo{journal}{Phys. Rev. Lett.} \textbf{\bibinfo{volume}{89}},
  \bibinfo{pages}{037903} (\bibinfo{year}{2002}).

\bibitem[{\citenamefont{Anders and Vedral}(2007)}]{Anders2007}
\bibinfo{author}{\bibfnamefont{J.}~\bibnamefont{Anders}} \bibnamefont{and}
  \bibinfo{author}{\bibfnamefont{V.}~\bibnamefont{Vedral}},
  \bibinfo{journal}{Open Syst. Inf. Dyn.} \textbf{\bibinfo{volume}{14}},
  \bibinfo{pages}{1} (\bibinfo{year}{2007}), \eprint{arXiv:quant-ph/0610268}.

\bibitem[{\citenamefont{Brandao and Plenio}(2008)}]{Brandao2008}
\bibinfo{author}{\bibfnamefont{F.~G. S.~L.} \bibnamefont{Brandao}}
  \bibnamefont{and} \bibinfo{author}{\bibfnamefont{M.~B.}
  \bibnamefont{Plenio}}, \bibinfo{journal}{Nature Physics}
  \textbf{\bibinfo{volume}{4}}, \bibinfo{pages}{873} (\bibinfo{year}{2008}).

\bibitem[{\citenamefont{Amico et~al.}(2008)\citenamefont{Amico, Fazio,
  Osterloh, and Vedral}}]{Amico2008}
\bibinfo{author}{\bibfnamefont{L.}~\bibnamefont{Amico}},
  \bibinfo{author}{\bibfnamefont{R.}~\bibnamefont{Fazio}},
  \bibinfo{author}{\bibfnamefont{A.}~\bibnamefont{Osterloh}}, \bibnamefont{and}
  \bibinfo{author}{\bibfnamefont{V.}~\bibnamefont{Vedral}},
  \bibinfo{journal}{Rev. Mod. Phys.} \textbf{\bibinfo{volume}{80}},
  \bibinfo{pages}{517} (\bibinfo{year}{2008}),
  \eprint{arXiv:quant-ph/0703044v3}.

\bibitem[{\citenamefont{Barvinsky and Kamenshchik}(2006)}]{Barvinsky2006}
\bibinfo{author}{\bibfnamefont{A.~O.} \bibnamefont{Barvinsky}}
  \bibnamefont{and} \bibinfo{author}{\bibfnamefont{A.~Y.}
  \bibnamefont{Kamenshchik}}, \bibinfo{journal}{JCAP}
  \textbf{\bibinfo{volume}{0609}}, \bibinfo{pages}{014} (\bibinfo{year}{2006}),
  \eprint{hep-th/0605132}.

\bibitem[{\citenamefont{Barvinsky and Kamenshchik}(2007)}]{Barvinsky2007a}
\bibinfo{author}{\bibfnamefont{A.~O.} \bibnamefont{Barvinsky}}
  \bibnamefont{and} \bibinfo{author}{\bibfnamefont{A.~Y.}
  \bibnamefont{Kamenshchik}}, \bibinfo{journal}{J. Phys. A}
  \textbf{\bibinfo{volume}{40}}, \bibinfo{pages}{7043} (\bibinfo{year}{2007}),
  \eprint{hep-th/0701201}.

\bibitem[{\citenamefont{Barvinsky}(2007)}]{Barvinsky2007b}
\bibinfo{author}{\bibfnamefont{A.~O.} \bibnamefont{Barvinsky}},
  \bibinfo{journal}{Phys. Rev. Lett.} \textbf{\bibinfo{volume}{99}},
  \bibinfo{pages}{071301} (\bibinfo{year}{2007}), \eprint{0704.0083}.

\bibitem[{\citenamefont{Hawking}(1984{\natexlab{b}})}]{Hawking1984}
\bibinfo{author}{\bibfnamefont{S.~W.} \bibnamefont{Hawking}},
  \bibinfo{journal}{Nucl. Phys. B} \textbf{\bibinfo{volume}{239}},
  \bibinfo{pages}{257} (\bibinfo{year}{1984}{\natexlab{b}}).

\bibitem[{\citenamefont{Kiefer}(2007)}]{Kiefer2007}
\bibinfo{author}{\bibfnamefont{C.}~\bibnamefont{Kiefer}},
  \emph{\bibinfo{title}{Quantum gravity}} (\bibinfo{publisher}{Oxford
  University Press, Oxford, UK}, \bibinfo{year}{2007}).

\bibitem[{\citenamefont{Alonso-Serrano
  et~al.}(2013{\natexlab{a}})\citenamefont{Alonso-Serrano, Garay, and
  Robles-P{\'e}rez}}]{Alonso2013}
\bibinfo{author}{\bibfnamefont{A.}~\bibnamefont{Alonso-Serrano}},
  \bibinfo{author}{\bibfnamefont{I.}~\bibnamefont{Garay}}, \bibnamefont{and}
  \bibinfo{author}{\bibfnamefont{S.}~\bibnamefont{Robles-P{\'e}rez}},
  \bibinfo{journal}{In preparation}  (\bibinfo{year}{2013}{\natexlab{a}}).

\bibitem[{\citenamefont{Strominger}(1990)}]{Strominger1990}
\bibinfo{author}{\bibfnamefont{A.}~\bibnamefont{Strominger}}, in
  \emph{\bibinfo{booktitle}{Quantum Cosmology and Baby Universes}}, edited by
  \bibinfo{editor}{\bibfnamefont{S.}~\bibnamefont{Coleman}},
  \bibinfo{editor}{\bibfnamefont{J.~B.} \bibnamefont{Hartle}},
  \bibinfo{editor}{\bibfnamefont{T.}~\bibnamefont{Piran}}, \bibnamefont{and}
  \bibinfo{editor}{\bibfnamefont{S.}~\bibnamefont{Weinberg}}
  (\bibinfo{publisher}{World Scientific, London, UK}, \bibinfo{year}{1990}),
  vol.~\bibinfo{volume}{7}.

\bibitem[{\citenamefont{Linde}(1988)}]{Linde1988}
\bibinfo{author}{\bibfnamefont{A.~D.} \bibnamefont{Linde}},
  \bibinfo{journal}{Phys. Lett. B} \textbf{\bibinfo{volume}{200}},
  \bibinfo{pages}{272} (\bibinfo{year}{1988}).

\bibitem[{\citenamefont{Robles-P{\'e}rez}(2012{\natexlab{b}})}]{RP2012}
\bibinfo{author}{\bibfnamefont{S.}~\bibnamefont{Robles-P{\'e}rez}}
  (\bibinfo{year}{2012}{\natexlab{b}}), \eprint{arXiv:1203.5774}.

\bibitem[{\citenamefont{Garay and Robles-P{\'e}rez}(2013)}]{Garay2013}
\bibinfo{author}{\bibfnamefont{I.}~\bibnamefont{Garay}} \bibnamefont{and}
  \bibinfo{author}{\bibfnamefont{S.}~\bibnamefont{Robles-P{\'e}rez}},
  \bibinfo{journal}{Submitted}  (\bibinfo{year}{2013}).

\bibitem[{\citenamefont{Birrell and Davies}(1982)}]{Birrell1982}
\bibinfo{author}{\bibfnamefont{N.~D.} \bibnamefont{Birrell}} \bibnamefont{and}
  \bibinfo{author}{\bibfnamefont{P.~C.~W.} \bibnamefont{Davies}},
  \emph{\bibinfo{title}{Quantum fields in curved space}}
  (\bibinfo{publisher}{Cambridge University Press, Cambridge, UK},
  \bibinfo{year}{1982}).

\bibitem[{\citenamefont{Mukhanov and Winitzki}(2007)}]{Mukhanov2007}
\bibinfo{author}{\bibfnamefont{V.~F.} \bibnamefont{Mukhanov}} \bibnamefont{and}
  \bibinfo{author}{\bibfnamefont{S.}~\bibnamefont{Winitzki}},
  \emph{\bibinfo{title}{Quantum Effects in Gravity}}
  (\bibinfo{publisher}{Cambridge University Press, Cambridge, UK},
  \bibinfo{year}{2007}).

\bibitem[{\citenamefont{Robles-P{\'e}rez}(2012{\natexlab{c}})}]{RP2012d}
\bibinfo{author}{\bibfnamefont{S.~J.} \bibnamefont{Robles-P{\'e}rez}},
  \emph{\bibinfo{title}{Inter-universal entanglement}}
  (\bibinfo{publisher}{Intech}, \bibinfo{year}{2012}{\natexlab{c}}),
  chap.~\bibinfo{chapter}{8}, \eprint{arXiv:1211.6366}.

\bibitem[{\citenamefont{Alonso-Serrano
  et~al.}(2013{\natexlab{b}})\citenamefont{Alonso-Serrano, Bastos, Bertolami,
  and Robles-P{\'e}rez}}]{Alonso2012}
\bibinfo{author}{\bibfnamefont{A.}~\bibnamefont{Alonso-Serrano}},
  \bibinfo{author}{\bibfnamefont{C.}~\bibnamefont{Bastos}},
  \bibinfo{author}{\bibfnamefont{O.}~\bibnamefont{Bertolami}},
  \bibnamefont{and}
  \bibinfo{author}{\bibfnamefont{S.}~\bibnamefont{Robles-P{\'e}rez}},
  \bibinfo{journal}{Phys. Lett. B} \textbf{\bibinfo{volume}{719}},
  \bibinfo{pages}{200} (\bibinfo{year}{2013}{\natexlab{b}}),
  \eprint{arXiv:1207.6852}.

\bibitem[{\citenamefont{Abramovitz and Stegun}(1972)}]{Abramovitz1972}
\bibinfo{editor}{\bibfnamefont{M.}~\bibnamefont{Abramovitz}} \bibnamefont{and}
  \bibinfo{editor}{\bibfnamefont{I.~A.} \bibnamefont{Stegun}}, eds.,
  \emph{\bibinfo{title}{Handbook of Mathematical Functions}}
  (\bibinfo{publisher}{NBS}, \bibinfo{year}{1972}).

\bibitem[{\citenamefont{Halliwell}(1987)}]{Halliwell1987}
\bibinfo{author}{\bibfnamefont{J.~J.} \bibnamefont{Halliwell}},
  \bibinfo{journal}{Phys. Rev. D} \textbf{\bibinfo{volume}{36}},
  \bibinfo{pages}{3626} (\bibinfo{year}{1987}).

\bibitem[{\citenamefont{Halliwell}(1990)}]{Halliwell1990}
\bibinfo{author}{\bibfnamefont{J.~J.} \bibnamefont{Halliwell}}, in
  \emph{\bibinfo{booktitle}{Quantum Cosmology and Baby Universes}}, edited by
  \bibinfo{editor}{\bibfnamefont{S.}~\bibnamefont{Coleman}},
  \bibinfo{editor}{\bibfnamefont{J.~B.} \bibnamefont{Hartle}},
  \bibinfo{editor}{\bibfnamefont{T.}~\bibnamefont{Piran}}, \bibnamefont{and}
  \bibinfo{editor}{\bibfnamefont{S.}~\bibnamefont{Weinberg}}
  (\bibinfo{publisher}{World Scientific, London, UK}, \bibinfo{year}{1990}),
  vol.~\bibinfo{volume}{7}.

\bibitem[{\citenamefont{Hawking}(1985)}]{Hawking1985}
\bibinfo{author}{\bibfnamefont{S.~W.} \bibnamefont{Hawking}},
  \bibinfo{journal}{Phys. Rev. D} \textbf{\bibinfo{volume}{32}},
  \bibinfo{pages}{2489} (\bibinfo{year}{1985}).

\bibitem[{\citenamefont{Robles-P{\'e}rez and
  Gonz{\'a}lez-D{\'\i}az}(2010)}]{RP2010}
\bibinfo{author}{\bibfnamefont{S.}~\bibnamefont{Robles-P{\'e}rez}}
  \bibnamefont{and} \bibinfo{author}{\bibfnamefont{P.~F.}
  \bibnamefont{Gonz{\'a}lez-D{\'\i}az}}, \bibinfo{journal}{Phys. Rev. D}
  \textbf{\bibinfo{volume}{81}}, \bibinfo{pages}{083529}
  (\bibinfo{year}{2010}), \eprint{arXiv:1005.2147v1}.

\bibitem[{\citenamefont{Kim and Page}(2001)}]{Kim2001}
\bibinfo{author}{\bibfnamefont{S.~P.} \bibnamefont{Kim}} \bibnamefont{and}
  \bibinfo{author}{\bibfnamefont{D.~N.} \bibnamefont{Page}},
  \bibinfo{journal}{Phys. Rev. A} \textbf{\bibinfo{volume}{64}},
  \bibinfo{pages}{012104} (\bibinfo{year}{2001}).

\bibitem[{\citenamefont{Lewis and Riesenfeld}(1969)}]{Lewis1969}
\bibinfo{author}{\bibfnamefont{H.~R.} \bibnamefont{Lewis}} \bibnamefont{and}
  \bibinfo{author}{\bibfnamefont{W.~B.} \bibnamefont{Riesenfeld}},
  \bibinfo{journal}{J. Math. Phys.} \textbf{\bibinfo{volume}{10}},
  \bibinfo{pages}{1458} (\bibinfo{year}{1969}).

\end{thebibliography}

\end{document}